\documentclass[preprint,10pt,plainnat]{aastex6}
\usepackage{amsmath,amstext}
\usepackage{psbox}
\usepackage{natbib}
\usepackage{graphicx}
\usepackage{latexsym}
\usepackage{amssymb,amsbsy}
\usepackage{eucal}
\usepackage{longtable}

\def\apjs{{\it Astrophys.~J.~Suppl.}}
\def\apj{{\it Astrophys.~J.}}

\def\apjl{{\it Astrophys.~J.~Lett.}}

\def\aap{{\it Astron.~Astrophys.}}

\begin{document}

%\begin{frontmatter}

%\shorttitle{New indexing and surface temperature analysis of exoplanets}
 
%\title{Habitability indexing of exoplanets and search for Mars-like planets}
\title{Similarity indexing of exoplanets in search for potential habitability: application to Mars-like worlds.}
%\title{Quantitative Comparison Between Earth-like and Mars-like Worlds}

\author{J.~M.~Kashyap\altaffilmark{1}}
\author{S.~B. Gudennavar\altaffilmark{2}}

\affil{Department of Physics, Christ University, Bengaluru 560 029, India}
\author{Urmi Doshi\altaffilmark{3}}
\author{M. Safonova\altaffilmark{4}}

\affil{M.~P.~Birla Institute of Fundamental Research, Bangalore 560 001, India}

\altaffiltext{1}{kas7890.astro@gmail.com}
\altaffiltext{2}{shivappa.b.gudennavar@christuniversity.in}
\altaffiltext{3}{urmi84@gmail.com}
\altaffiltext{4}{margarita.safonova62@gmail.com}

\begin{abstract}
Study of exoplanets is one of the main goals of present research in planetary sciences and astrobiology.  Analysis of huge planetary data from space missions such as CoRoT and Kepler is directed ultimately at finding a planet similar to Earth -- the Earth's twin, and answering the question of potential exo-habitability. The Earth Similarity Index (ESI) is a first step in this quest, ranging from 1 (Earth) to 0 (totally dissimilar to Earth). It was defined for the four physical parameters of a planet: radius, density, escape velocity and surface temperature. The ESI is further sub-divided into interior ESI (geometrical mean of radius and density) and surface ESI (geometrical mean of escape velocity and surface temperature). The challenge here is to determine which exoplanet parameter(s) is important in finding this similarity; how exactly the individual parameters entering the interior ESI and surface ESI are contributing to the global ESI. Since the surface temperature entering surface ESI is a non-observable quantity, it is difficult to determine its value. Using the known data for the Solar System objects, we established the calibration relation between surface and equilibrium temperatures to devise an effective way to estimate the value of the surface temperature of exoplanets for further analysis with our graphic methodology.

ESI is a first step in determining potential exo-habitability that may not be very similar to a terrestrial life. A new approach, called Mars Similarity Index (MSI), is introduced to identify planets that may be habitable to the extreme forms of life. MSI is defined in the range between 1 (present Mars) and 0 (dissimilar to present Mars) and uses the same physical parameters as ESI. We are interested in Mars-like planets to search for planets that may host the extreme life forms, such as the ones living in extreme environments on Earth; for example, methane on Mars may be a product of the methane-specific extremophile life form metabolism. 

\end{abstract}

\keywords{Earth Similarity Index (ESI), Mars Similarity Index (MSI), potential habitability, surface temperature, exoplanets}

%\end{frontmatter}
%\maketitle

\section{Introduction}

The search for life elsewhere outside the Earth has been a very fascinating area in recent years. A lot of efforts have been channeled in this direction in the form of space missions looking for potential habitable planets \citep[e.g.][]{sea}. Analysis of huge volume of collected planetary data from space missions such as CoRoT and Kepler is directed at finding a planet similar to Earth -- the Earth's twin, and answering the question of exo-habitability. A full assessment of the habitability of any planet requires very detailed information about it, which we still cannot achieve. Currently, the best we can do is to compare the properties we can measure (or infer), in other words determine the potential habitability. Here, we consider habitability in the Earth way, or in some modified but still recognizable by us way (e.g. mars, or Titan, or Europa/Enciladus). The key parameters that are typically available from observations are stellar flux that can help to judge the temperature of the planet, and the planet's size, derived either from transit observation -- radius, or from RV measurements -- low limit on mass. In the latter case, from the statistical point of view the true mass differs from the low limit by only a factor of $4/\pi \sim 1.2$ (e.g. Maruyana et al. 2013). With the developed theoretical models, it is already possible to exclude planets from the potential habitability list based on mass/radius values. Thus, since we prefer to start the search only on the rocky worlds, only planets of up to $1.6\,R_{\rm E}$ or $2\,M_{\rm E}$ can be rocky (e.g. Grosset, Mocqut\&Sotin, 2007; Chen \& Kipping, 2016). With the basic criteria and data already available, we can start gauging the potential habitability of exoplanets. The challenge here is to determine which exoplanet parameter(s) are important in finding this similarity. To address this challenge, the Earth Similarity Index (ESI), a parametric index to analyze the exoplanets’ data, was introduced to assess the Earth-likeness of exoplanets \citep{schu,mascaro}. This multi-parameter ESI scale depends on radius, density, escape velocity and surface temperature of a planet. The ESI index is used by the Planetary Habitability Laboratory (PHL), University of Puerto Rico, Arecibo, to estimate the potential habitability of all discovered to date exoplanets. Here, the total ESI ranges from 0 (totally dissimilar to Earth) to 1 (identical to Earth). A planet with ESI $\geq 0.8$ is considered an Earth-like, but even planets with ESI $\geq 0.73$ (e.g. Mars) are optimistically called potentially habitable planets (PHPs). As of now (January 2017), there are 44 assumed PHPs\footnote{http://phl.upr.edu/projects/habitable-exoplanets-catalog}. We focus on indexing the exoplanet data taken from the online Exoplanets Catalog (PHL-EC)\footnote{http://phl.upr.edu/projects/habitable-exoplanets-catalog/data/database} maintained by the PHL. The ESI is subdivided into the interior ESI --- an estimate of
probability of a planet to have a rocky interior (calculated from radius and density), and exterior ESI -- a surface similarity, an estimate of probability of a surface temperature to be within a habitable range (calculated from surface temperature and escape velocity); the total (global) ESI is their geometric mean. This way, the scale is useful for the overall concept of planet similarity for both interior and surface properties. One of our objectives here is to establish how the individual parameters, entering the interior and surface ESI, are contributing to the global ESI, using graphic analysis approach.  

Presently, one of the ESI parameters, the mean surface temperature, is estimated for the rocky planets in PHL-EC by following a correction factor of 30-33 K based on the Earth's greenhouse effect. Another  objective of our work in calculations of the global ESI is to try to introduce a better estimate of the surface temperature of rocky exoplanets, because it may be a crucial element in the search for habitable planets (Carone et al. 2016). From graphic analysis of the known data for some of the Solar System objects, we establish the calibrated relation between surface and equilibrium temperatures. This relation is extended to estimate the surface temperatures of exoplanets from their equilibrium temperatures.

Though Earth is currently the only place where life as we know it exists, there is a good probability that life could have existed on Mars in the past (Abramov \& Mojzsis, 2016). Indeed, Mars is technically inside the Sun's habitable zone (HZ),  though its ESI is only 0.73 (PHL). However, Mars-like conditions may be suitable for the extremophile life forms, such as the ones living in extreme environments on Earth (numerous experiments have proven several organisms to survive the simulated martian conditions, e.g. Onofri et al. 2015). Reports on the presence of methane in the martian atmosphere (Webster et al. 2015)
raise the possibility of the existence of a methane-specific extremophile life forms; in fact, on Earth methanogens thrive in conditions similar to martian environments: in dry desert soils and in 3-km deep glacial ice (see Hu et al. 2016, and references therein). Therefore, we are interested in Mars similarity to look for planets potentially habitable by extremophiles, say extreme PHPs. For that, we introduce the new indexing formulation, called Mars Similarity Index (MSI), where the range is 0 (no similarity to present Mars) and 1 (present Mars). Mars here represents a sort of test-bed that can indicate the potential habitability of small rocky planets. A planet with MSI $> 0.63$ is recognized to be Mars-like, e.g. Earth, which is a habitable planet. We find from our data analysis six Mars-like planets (Mars, Earth, Kepler-186~f, Kepler-442~b, Kepler-438~b, Kepler-62~f, GJ 667C~c, GJ 667C~f), which are also included in the PHP list. 

The structure of this paper is as follows. In Section~2, we give the detailed derivation of the ESI and MSI scales. Data analysis is presented in Section~3, which includes ESI analysis, surface temperature analysis, and MSI analysis as subsections. Finally, discussion and conclusions are given in Section~4.

\section{Mathematical formulation of the ESI and MSI}

Distance/similarity measurements are widely used in, for example, classification of objects, clustering and retrieval problems, or to compute the overlaps between quantitative data. Here, the distance $d$ is represented as dissimilarity, and proximity is equivalent to similarity $s$. Mathematically, the concept of distance is a metric one -- a measure of a true distance in Euclidean space ${\bf R}^n$. This problem can be addressed by using Minkowski's space of $L_p$ form \citep{Sung}, in which $p$-norm stands for finite $n$-dimensional vector space,
\begin{equation}
d_{\rm } =\left( \sum_{i=1}^{n}|p_i-q_i|^p \right)^{1/p}\,,
\label{eq:Lp}
\end{equation}
where the city-block, or Manhattan, $L_1$ distance 
\begin{equation}
d_{\rm CB} = \sum_{i=1}^{n}|p_i-q_i| \,,
\label{SIIeq1}
\end{equation}
and Euclidean $L_2$ distance
\begin{equation}
d_{\rm Euc} =\left( \sum_{i=1}^{n}(p_i-q_i)^2\right)^{1/2}\,,
\label{eq:euc}
\end{equation}
are the special cases. Here, $p_i$ and $q_i$ are the coordinates of $p$ and $q$ in dimension $i$, and $i= 1, 2, 3,\ldots, n$. $L_1$ form has an advantage that it can be decomposed into contributions made by each variable being the sum of absolute differences (e.g. for the $L_2$ form, it would be the decomposition of the squared distance). However, in comparing multivariate data sets, some distance measures can be applied that do not conform to the usual definition of a metric (i.e. metric axioms, e.g. Deza and Deza 2016), but are still very useful as a measure of difference (or similarity) between samples. Here, we are interested in finding similarities between different planets based on their various characteristics, in other words, a multivariate data sets. In such cases, especially abundant in ecological and environmental studies to quantify the differences between samples collected at different sampling locations \citep[e.g][]{12}, the Bray-Curtis distance is the most used scale \citep{BrayCurtis}; it is also sometimes called an ecological distance \citep{TreeBook}. The advantages in using the ecological distance is that differences between datasets can be expressed by a single statistic. Bray-Curtis is a modified Manhattan distance\footnote{Bray-Curtis distance becomes Manhattan when it is applied to relative counts, as opposed to the absolute abundances \citep{Book}.}, where the summed differences between the variables are standardized by the summed variables of the objects,
\begin{equation}
d_{\rm BC} = {\sum_{i=1}^{n}|p_i-q_i|  \over \sum_{i=1}^{n}\left(p_i+q_i\right)}\,.
\label{SIIeq2}
\end{equation}
Here $p_i$ and $q_i$ are two different precisely measurable quantities between which the distance is to be measured, and $n$ is the total number of variables. The assumption in the Bray-Curtis scale is that samples are taken from same physical  measure, e.g. mass, or volume. It is because the distance is found from the raw counts, so that if there is a higher abundance in one sample comparing to the other, it is a part of the difference between the two samples. The advantage of the Bray-Curtis scale is the simplicity in interpretation: $0$ means the samples are exactly the same, and $1$ means they are completely disjoint. It shall be kept in mind that Bray-Curtis distance is not the true metric distance since it violates the triangular inequality -- one of the axioms of a true metric. Though it is still called a semi-metric in some earlier works, in modern mathematics it is defined as a non-metric \citep{Deza}.

The intersection between two distributions is a more widely used form of similarity (as opposed to distance). Most similarity measures for intersection can be transformed from the distance measure by the transformation technique (see Bloom 1981, but not exclusively),
\begin{equation}
s_{\rm BC} = 1- d_{\rm BC} =
1- {\sum_{i=1}^{n}|p_i-q_i|  \over \sum_{i=1}^{n}\left(p_i+q_i\right)}\,.
\label{eq:similarity}
\end{equation}
Here, the value of $0$ means complete absence of relationships, and the value of $1$ shows a complete matching of the two data records in the $n$-dimensional space \citep{JanSchulz}.

Distances/similarities based on heterogeneous data can be found after a process of standardization --- balancing of the contribution of different types of variables in an equitable way \citep{Book}. One way to do that is to calculate the similarity for each set of homogeneous variables and then combine them using various methods (see Sec.~\ref{sec:ESI CALCULATION}). Higher values in one set may impact the result of the Bray-Curtis similarity more dominantly and imply that these variables are more likely to discriminate between sets. Therefore, user-defined weighting is a convenient (though subjective) method for down-weighing the differences for a set of variables.  

In our case, we would like to compare sets of different variables of one planet with that of the reference value, for example, Earth, to find  planets that are similar to Earth. We rewrite  Eq.~(\ref{eq:similarity}) as
\begin{equation}
s=\left[1- \frac{|x-x_0|}{(x+x_0)}\right]^{w_x}\,,
\label{last_sim}
\end{equation}
where $x$ is the physical property of the exoplanet, $w_x$ is the weight for this property, $x_0$ is the reference value, and the dimension $n=1$, since we are constructing the index separately for each physical property. We find the weights by defining the threshold value ($V$) in the similarity scale for each quantity,
\begin{equation}
V = \left[1-\Big|\frac{x_0-x}{x_0+x}\Big| \right]^{w_x}\,.
\label{SIIeq6}
\end{equation}
Traditionally, the similarity indices are subdivided into equal 0.2 intervals \citep{Bloom}, defining very low, low, moderate, high and very high similarity regions. Therefore, the threshold can be defined on this grounds, for example considering only very high similarity region with the  threshold $V=0.8$. Defining the physical limits $x_a$ and $x_b$ of the permissible variation of a variable with respect to $x_0$ (i.e. $x_a<x_0<x_b$), we calculate the weight exponents for the lower $w_a$ and upper $w_a$ limits,
\begin{equation}
w_a = \frac
{\ln{V}}
{\ln\left[1-\left|\frac{x_0-x_a}
{x_0+x_a}\right|\right]}
\,,\quad 
w_b = \frac
{\ln{V}}
{\ln\left[1-\left|\frac{x_b-x_0}{x_b+x_0}\right|\right]}
\,,\quad 
\label{eq:weight_exponent}
\end{equation}
The average weight is found by the geometric mean,
 \begin{equation}
w_x=\sqrt{{w_a}\times {w_b}}\,.
\label{eq:geom_mean}
\end{equation}

In this paper, we use Eq.~\ref{eq:similarity} to define the Earth similarity index 
\begin{equation}
ESI_x = {\left[1-\Big|
\frac{x-x_0}{x+x_0}\Big| \right]^{w_x}}\,,
\label{eq:esi}
\end{equation}
and Mars similarity index
\begin{equation}
MSI_x = {\left[1-\Big|
\frac{x-x_0}{x+x_0}\Big| \right]^{w_x}}\,,
\label{msi}
\end{equation}
where $x$ is the physical property of the exoplanet (for example, radius or density), and $x_0$ is the reference to Earth in ESI, and to Mars in MSI.
 
We focus on indexing the exoplanet data taken from the online PHL Exoplanets Catalog (PHL-EC)\footnote{Maintained by the PHL, http://phl.upr.edu/projects/habitable-exoplanets-catalog/data/database}. The catalog contains more than 60 observed and derived stellar and planetary parameters for all currently confirmed exoplanets.
The input parameters for similarity scales are radius $R$, density $\rho$, surface temperature $T_S$ and escape velocity ${V_{e}}$. These parameters, except the surface temperature, are expressed in Earth Units (EU) in the ESI calculations. In addition, we normalized the mean radius, bulk density and escape velocity to Mars Units (MU) in the calculations of the Mars Similarity Index. The corresponding weight exponents for both ESI and MSI scales were found using the threshold value $V= 0.8$, indicating very high similarity region. The weight exponents for the upper and lower limits of parameters were calculated for the Earth-like parameter range \citep{schu}: radius 0.5 to 1.9 EU, mass 0.1 to 10 EU, density 0.7 to 1.5 EU, surface temperature 273 to 323 K, and escape velocity 0.4 to 1.4 EU, through Eqs.~(\ref{eq:weight_exponent}) and (\ref{eq:geom_mean}). Similarly, the weight exponents for the lower and upper limits of parameters were defined for the Mars-like conditions: radius range 0.72 to 1.88 MU, mass range 0.514 to 9.30 MU, density range 0.89 to 1.402 MU, surface temperature range 233 to 418 K, and escape velocity range 0.85 to 2.23 MU. Here, MU are Mars Units, where radius is 3390 km, density is 3.93 g/cm$^3$, escape velocity is 5.03 km/s, and the mean surface temperature 240 K \citep{MarsBook}. The reason behind the limits definitions were to have a rocky planet, with lower limit in comparison to Mars (mass and radius are chosen as for Mercury, the smallest planet in our Solar System, and density as for Io), and with Earth as the upper limit. The temperature range is chosen on the basis of the range known to be suitable for extremophile life forms on our planet, between $-40^{\circ}$ and $+145^{\circ}$C \citep{Tung2005}. The corresponding weight exponents were calculated using the same method as for the ESI, and are given in Table~\ref{table:3.1} along with Earth and Mars unit parameters.

\begin{table}[h!]
\begin{center}
\caption{ESI and MSI Parametric Table}\label{table:3.1}
\begin{tabular}{l*{4}{c}r}
\hline
Planetary Property &Ref. Value & Ref. Value & Weight Exponents & Weight Exponents\\
            & for ESI & for MSI & for ESI & for MSI\\ \hline
Mean Radius	& 1EU &	1MU & 0.57 & 0.86\\ 
Bulk Density &	1EU &1MU &	1.07 & 2.10 \\ 
Escape Velocity	&1EU&	1MU&	0.70 & 1.09\\ 
Surface Temperature	& 288K&	240K&5.58 & 3.23\\
\hline
\end{tabular}\\[0.1in]
{\small EU = Earth Units, with Earth radius 6371 km, density 5.51 g/cm$^3$, escape velocity 11.19 km/s. \\
MU = Mars Units, with Mars radius 3390 km, density 3.93 g/cm$^3$, escape velocity 5.03 km/s.}
\end{center}
\end{table}

\section{Data analysis}

The PHL-EC contains 3635 confirmed exoplanets (as of January 2017). However, some  of them do not have all the required input parameters for calculating the global ESI. Wherever we could, we supplemented the missing data after the additional search through the following catalogs: Habitable Zone Gallery, Open Exoplanet Catalogue, Extrasolar Planets
Encyclopaedia,  Exoplanets Data Explorer and Nasa Exoplanet Archive. In addition, we discarded the entries with unrealistic values. For example, some of the planets had the equilibrium/surface temperatures of less than 3.2 K, and some had the density values of around 500 EU. In such cases, we have done extensive search through all available exoplanet catalogs and discovery papers, and supplemented those values that we could find. As an example, there is a lot of confusion with the available data on Kepler-53c, Kepler-57c and Kepler-59b planets. Their densities are listed in the PHL-EC as 162, 573.18 and 492 Earth Units, respectively. In such cases, the density of, say, Kepler 57c, becomes 21 times the density of the Sun's core. There is obviously a mistake in the retrieved data. We have searched through the catalogs and found that for Kepler 53c, the mass of 5007.56 EU used in calculating that density was, in fact, an upper limit from the stability analysis (Steffen et al. 2013), and it was subsequently updated (Haden et al. 2014) to $35.4^{+19.}_{-14.8}$ EU with nearly the same value for the radius. Using this number, the density becomes 1.169 EU. Similarly, for Kepler 57c, the density of 573.18 EU was obtained using the upper limit on mass of 2208.83 EU which, after updating the mass to $7.4^{+9.4}_{-6.3}$ EU (Haden et al. 2014) with essentially the same radius, fell within the normal range, 1.139 EU. Correspondingly, we have corrected the data for these planets in our catalog. Some of the entries had to be removed owing to the absence of available data (or very conflicting values), which left us with 3566 exoplanets for our analysis (1650 of them are rocky).

\subsection{SURFACE TEMPERATURE ANALYSIS}

Surface temperature of the planet enters the exterior ESI index -- a measure of surface similarity. Usually, the extrasolar planets temperatures are estimated from the calculated temperature of the parent star and other observational data (e.g. distance to the star, etc.) \citep{Nature}, in the assumption that the planet does not have an atmosphere e.g.  \citep[e.g.][]{williams}, which gives a so-called radiative equilibrium temperature. It is essentially an effective temperature attained by an isothermal planet that is in complete radiation equilibrium with the parent star. If a planet has no atmosphere, this temperature equals the surface temperature of the planet. Actual temperatures of the gas giants are usually higher because there is an internal heat source. If the planet has a substantial atmosphere, then this is the temperature near the tropopause of the planet -- approximately the level where the radiation from the planet is emitted to space. Terrestrial planets with atmospheres, such as Earth or Venus, for example, have higher temperatures due to the greenhouse effect. So, though equilibrium temperature may be different from the actual surface temperature of the planet, it is a useful parameter in comparing planets. It has been used to estimate the boundaries of Habitable Zones around stars (e.g. Kaltenegger and Sasselov 2011; Schulze-Makuch et al. 2011), and is an essential quantity for exoplanets when it is not possible to measure (or calculate) the surface T. 

\begin{figure}[h!]
\centering
\includegraphics[width=9cm,angle=0]{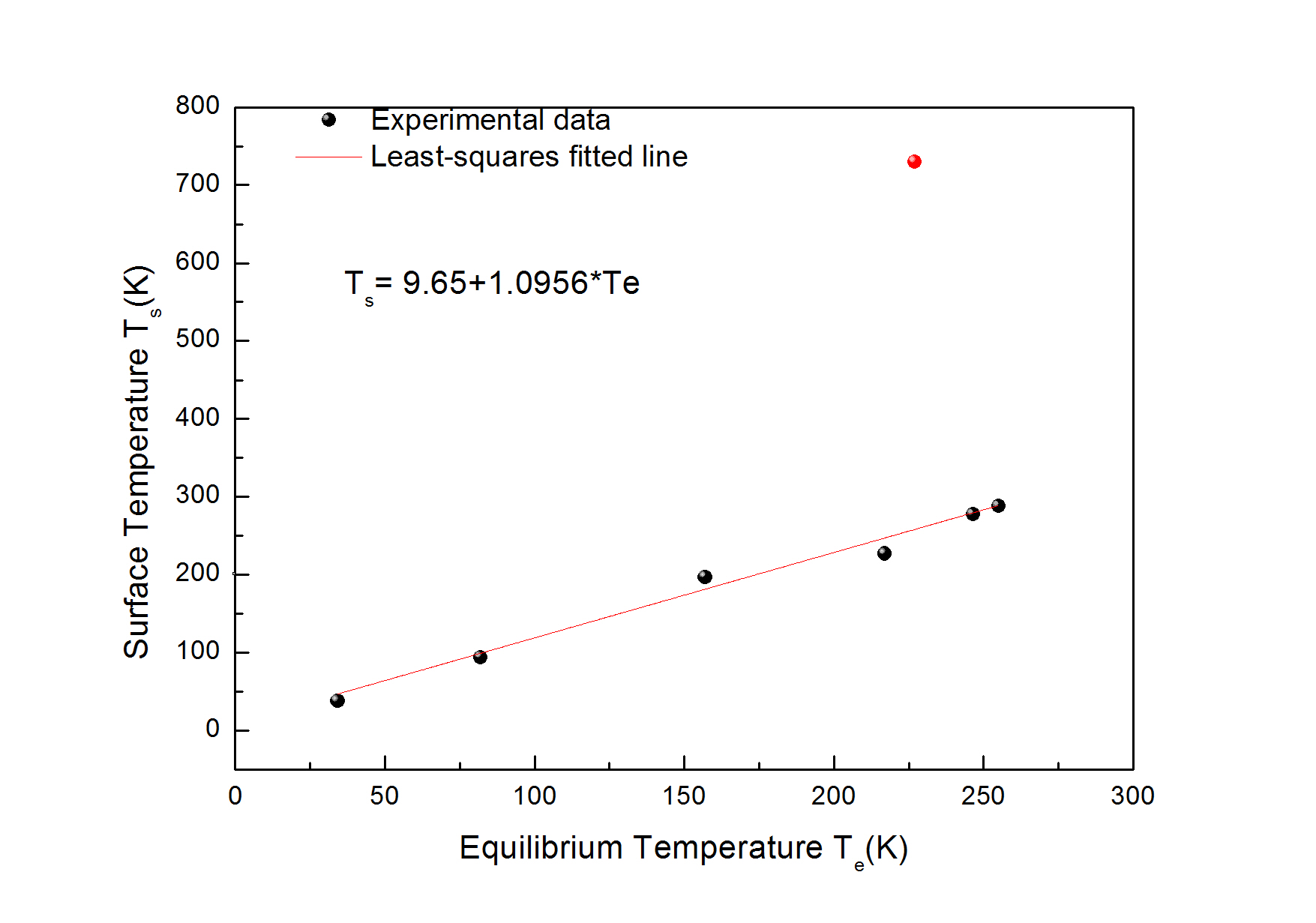} 
\caption{Calibration of surface temperature. Venus is masked as red dot due its very high surface temperature.\label{figure: 3.4}}
\end{figure}

The albedo entering the surface temperature equation is generally not known and has to be assumed. The albedo depends on many factors, some of which are geometry, composition,  and atmospheric properties. In addition, there is also the emissivity index which  represents atmospheric effects, and depends on the type of surface and many climate models. In calculations of the surface temperatures of rocky planets by the PHL, a correction factor of $30-33$ K is usually used, based on the Earth's green-house effect \citep{{schu},{Vol}}. However, we shall not ignore the good quality data available for other planets in our Solar System, because unlike with the exoplanets, many parameters are actually been measured. We have used the known data of our Solar System objects (only rocky bodies), using several available sources, such as NASA planetary fact sheet \citep{williams}, and extrapolated the resulting relation to obtain the surface temperatures of rocky exoplanets in our dataset. The relation between the surface temperature $(T_s)$ and the equilibrium temperature $(T_e)$ (Fig.~\ref{figure: 3.4}) is
\begin{equation}
T_s=9.65+1.0956\times T_e\,,
\label{eq:regression}
\end{equation}

In Table~\ref{Table:3.3}, we present the equilibrium and surface temperatures of several Solar System objects, and the sample of our results for a few potentially habitable exoplanets. 

\begin{table}[h!]
\centering
\caption{Equilibrium and surface temperatures}
\begin{tabular}{l*{4}{c}r}
\hline
Planet	& Equilibrium Temp &	Surface Temp &  $T_s=9.65+1.0966 \, T_e$\\
\ & (K)	&	(K)	& (K)\\ \hline
Earth	& 255	& 288			&	289.28\\
Mars	& 217	& 240			&	247.61\\
Moon	& 157	& 197			&	181.81\\
Venus	& 227	& 730			&	258.57\\
Titan	& 82	& 94			&	99.57\\
Triton	& 34.2	& 38			&	47.15\\
GJ 667Cc &	246.5 &	277.4		&	279.96\\
Kepler-442 b&	233	& unknown	&	265.16\\
Kepler-438 b&	276	& unknown	&	312.31\\
GJ 667 C f	& 220.7	& unknown	&	251.67\\
\hline
\end{tabular}
\label{Table:3.3}
\end{table}
Surface temperature of the Moon varies from Farside 120 K to the 350 K of a subsolar point, hence we adopted the average surface temperature for the Moon of 197 K as the most physically robust estimate \citep{Vol}. 

This empirical correlation is, of course, very simplified. However, such regression analysis was already successfully used to predict the mean global temperatures of rocky planets in \citet{Vol1} (see their Figs.~1 -- 4). The objective of our analysis was to evaluate how well the resulting new relation may predict the observed mean surface temperatures; in Table~\ref{Table:3.3} we show the measured and calculated by Eq.~12 surface temperatures, and observe that calculated values are pretty close to the measured ones. 

\subsection{ESI CALCULATION}
\label{sec:ESI CALCULATION}

First we converted the input parameters to Earth Units (EU), except the surface temperature, which is left in Kelvin. Some of the planets in the PHL-EC do not have estimates for the surface temperature. To mitigate this, we have calculated the surface temperature of all rocky  planets using Eq.~(\ref{eq:regression}). The corresponding Earth Similarity Index for each parameter was calculated using Eq.~(\ref{eq:esi}). Then these indices were  separately combined to form an interior similarity and surface similarity. The interior ESI is thus
\begin{equation}
ESI_I = \sqrt{{ESI_R} \times {ESI_\rho}}\,,
\label{eq:interiorESI}
\end{equation}
and surface ESI is:
\begin{equation}
ESI_S = \sqrt{ESI_T \times ESI_{V_e}}\,,
\label{eq:surfaceESI}
\end{equation}
where $ESI_R$, $ESI_\rho$, $ESI_T$ \&  $ESI_{V_e}$ are Earth Similarity Indices, calculated for radius, density, surface temperature and escape velocity, respectively. The global ESI is their geometric mean: 
\begin{equation}
ESI = \sqrt{{ESI_I} \times {ESI_S}}\,.
\label{eq:globalESI}
\end{equation}
The results of the ESI calculations for all 3566 currently confirmed exoplanets are presented in Table~\ref{Table:3.2} (only few planets are shown as an example). The full dataset is available online (Kashyap et al. 2017). The sample calculation of ESI for Mars is given in the Appendix. 

\begin{table}[h!]
\centering
\caption{A sample of calculated ESI}
\begin{tabular}{l*{8}{c}r}
\hline
Names & Radius & Density &Temp & E. Vel  & $ESI_S$ & $ESI_I$ & ESI\\ 
\ &  (EU) & (EU) & (K) & (EU) &  \ & \ &  \\\hline
Earth &	1.00 &	1.00 &	288	&1.00	&	1.00 &1.00	&1.00\\
Mars &	0.53 &	0.73 &	240	&0.45	&	0.65	&0.82	&0.73\\
Kepler-438 b & 1.12 & 0.90  & 312 & 1.06&  0.88 &0.95 & 0.91\\
Proxima Cen b & 1.12 & 0.90  & 259 & 1.06&  0.85 &0.95 & 0.90\\
GJ 667C c & 1.54 & 1.05 & 280 & 1.57& 0.88  & 0.92 & 0.90\\
Kepler-296 e & 1.48 & 1.03  & 302 & 1.5&  0.86 &0.93 & 0.89\\
\hline
\end{tabular}
\label{Table:3.2}
\end{table}

\subsection{ESI ANALYSIS}

\begin{figure}[h!]
\centering
\includegraphics[width=9cm,angle=0]{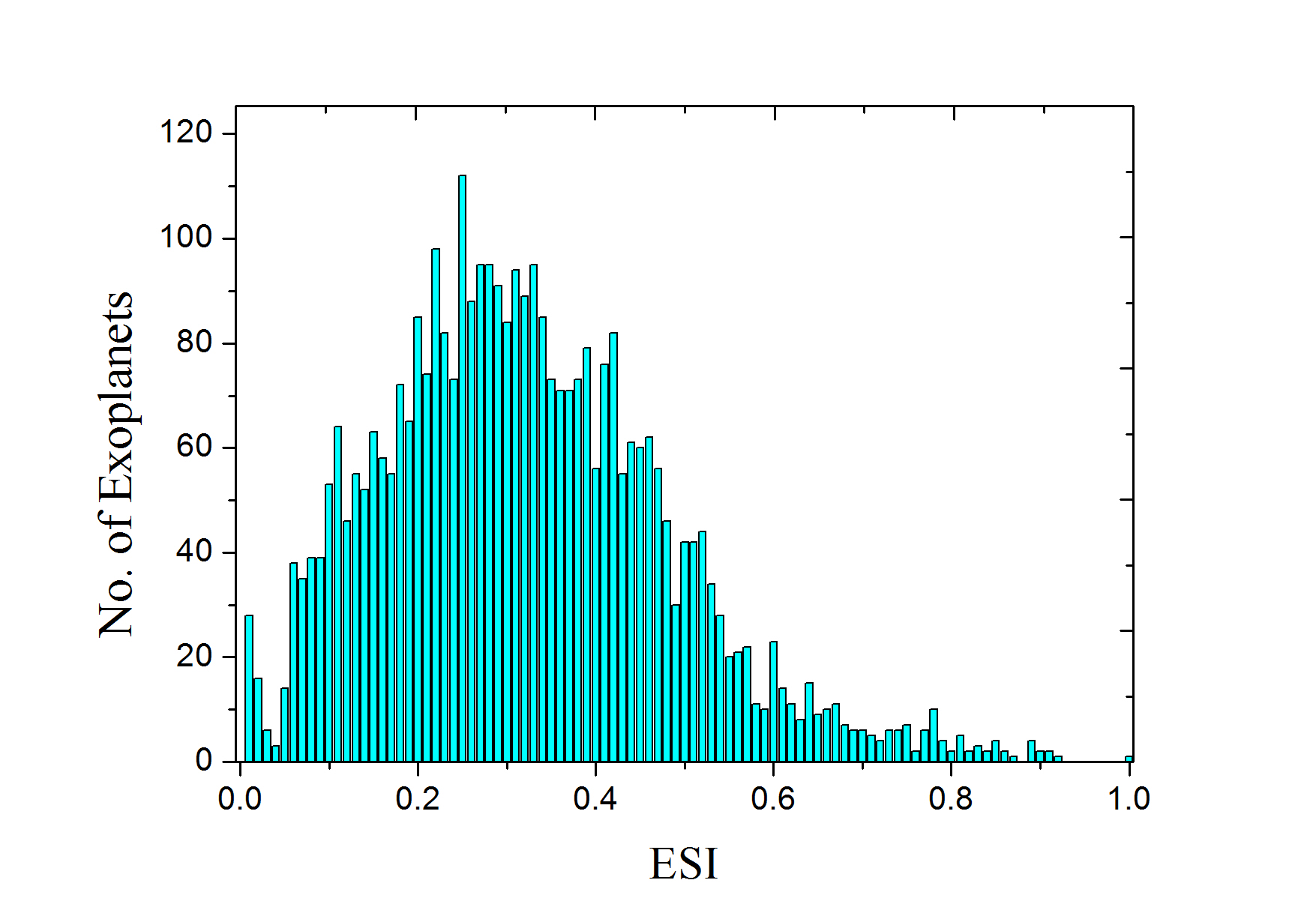} 
\caption {Histogram of the global ESI values of 3566 exoplanets} 
\label{figure: 3.2}
\end{figure}

\begin{figure}[h!]
\centering        
\includegraphics[width=9cm,angle=0]{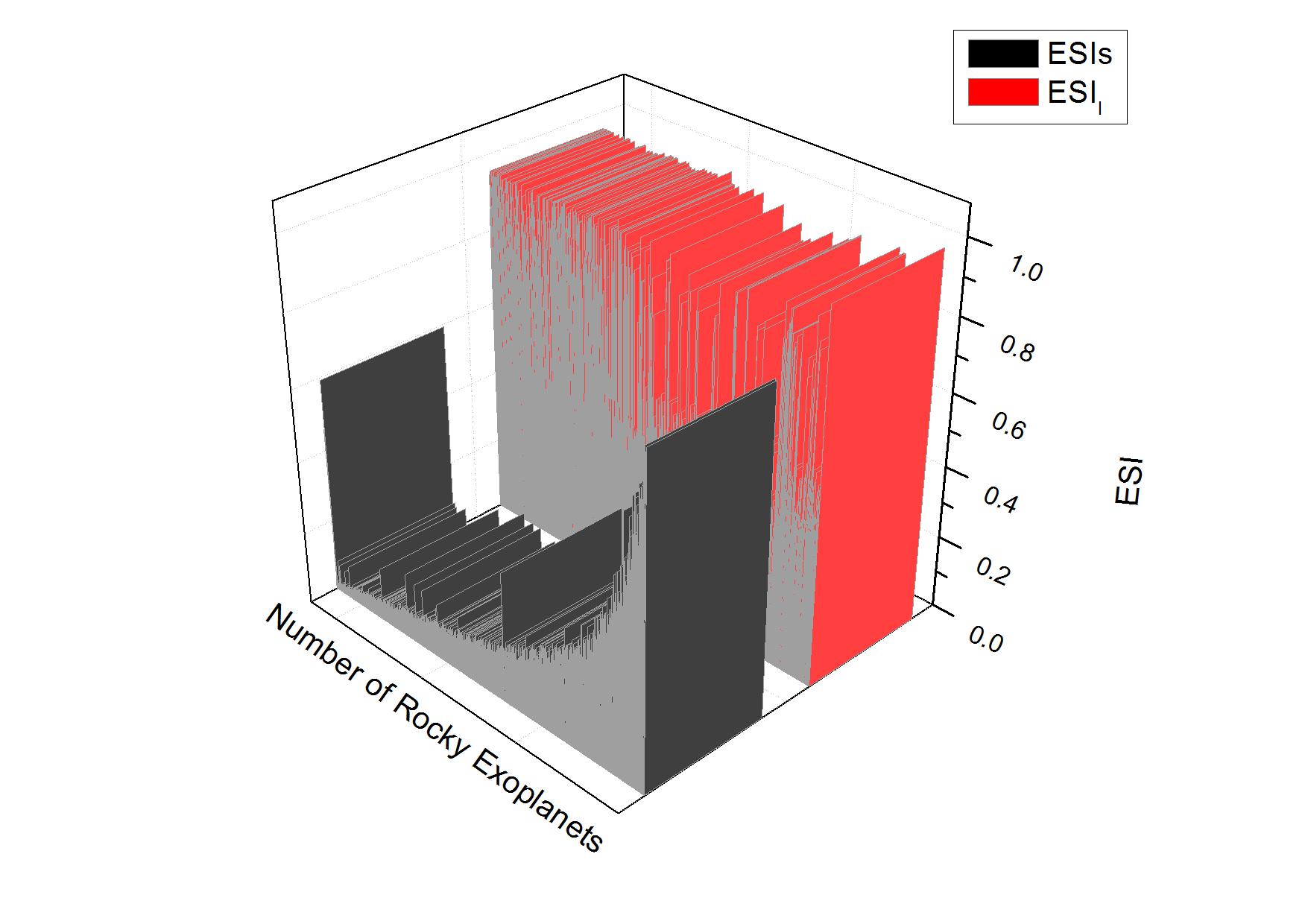} 
\caption{3-D histogram of interior and surface ESI for 1650 rocky planets.}
\label{figure: 3.1}
\end{figure}

Results of Section~3.1 are presented as a histogram of global ESI (Fig.~\ref{figure: 3.2}). According to the PHL project, surface ESI is dominating the interior ESI, because the surface temperature weight exponent value is much higher than that of the interior parameters. We found, however, that this is only true for the giant planets. For the rocky planets, we found that the interior ESI is a predominant factor in the global ESI, where the real values of interior and surface ESI play a larger role than the weight exponent. The 3-D histogram (Fig.~\ref{figure: 3.1}) is the result of overplotting interior and surface ESI for all the rocky exoplanets. 

\begin{figure}[h!]
\centering        
\includegraphics[width=13cm,angle=0]{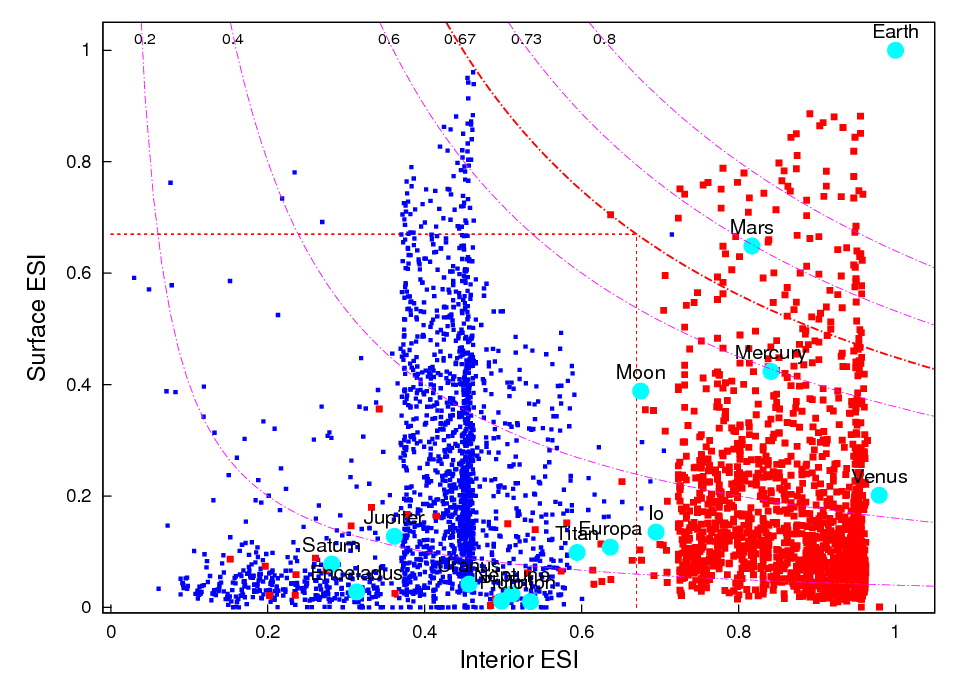} 
\caption{Plot of interior ESI versus surface ESI. Blue dots are the giant planets, red dots are the rocky planets, and cyan circles are the Solar System objects (Table~3). The dashed curves are the isolines of constant global ESI, with values shown in the plot. The giant planets above 0.67 dotted lines are of water-gas composition, and the planets to the right of 0.67 dotted line are rocky planets. Planets with ESI$\gtrsim 0.8$ are considered Earth-like. However, the optimistic limit is $\sim 0.67$.}
\label{fig:ESIplot}
\end{figure}

In Fig.~\ref{fig:ESIplot}, we present a scatter plot of interior ESI versus surface ESI. Blue dots are the giant planets, red dots are the rocky planets, and cyan circles are the Solar System objects. The dashed curves are the isolines of constant global ESI, with values shown in the plot. Planets above ESI=0.8 are conservatively considered Earth-like, and planets with ESI $\gtrsim 0.73$ are optimistically potentially habitable planets (PHL). We see that there are 29 Earth-like planets (ESI > 0.8) in 3566 planets that we have considered. In this plot we also see the predominant nature of the interior ESI. However, due to the geometrical mean nature of the global ESI formula, we need to consider all the four parameters to check the habitability of the planet.
We also find from the plot that there seem to be a definite division between gaseous and rocky planets, at approximately $ESI_I=0.67$ (interior ESI of the Moon). It is interesting to note that this division separates Moon and Io, rocky satellites, (especially Io, which is closer in bulk composition to the terrestrial planets) and, say Pluto and Europa, which are composed of water ice--rock.

When the ESI index was proposed, it was accepted that even planets with ESI between 0.6 and 0.8 could be potentially habitable, or at least similar to Earth. Thus, we propose to extend the optimistic limit from 0.73 to 0.67. There are now we about 100 rocky exoplanets with ESI $> 0.67$. For example, the ESI of Kepler-445 d is 0.76. It is located in the HZ, and has an estimated surface temperature of 305 K, which would make it suitable for life. 

\subsection{MSI CALCULATION}

In our Solar System, we have discovered many planetary bodies with conditions similar to some of the terrestrial environments where we know that Earth extremophiles live. Particularly notable here is Mars. It is now believed that Mars had a much wetter and warmer environment in its early history (Grotzinger et al. 2015; Wray et al. 2016). The discovery of a desert varnish on Mars (Krinsley et al. 2012), believed to be the product of the specific bacteria on Earth (Dorn and Oberlander 1981), has gotten us even further down the road of whether life existed/exists on Mars with its super-extreme conditions for habitation. We introduce here the Mars Similarity Index (MSI), calculated using Eq.~(\ref{msi}), to look for Mars-like planets as potential planets to host extremophile life forms. We are interested in this study to compare extreme environments similar to Mars.  
\noindent
Interior MSI is
 \begin{equation}
MSI_I = \sqrt{{MSI_R} \times {MSI_\rho}}\,,
\end{equation}
and the surface MSI is
\begin{equation}
MSI_S = \sqrt{MSI_{T_S} \times MSI_{V_e}}\,,
\end{equation}
where $MSI_R$, $MSI_\rho$, $MSI_{T_S}$ \&  $MSI_{V_{e}}$ are Mars similarity indices calculated for radius, density, surface temperature and escape velocity, respectively. The global MSI is given by:
 \begin{equation}
MSI = \sqrt{{MSI_I} \times {MSI_S}}
\end{equation}

A result of MSI calculations is  presented in Table~\ref{Table:msi} (only few entries are shown). The full dataset is available online (Kashyap et al. 2017). For the graphic analysis part we have used all 3566 confirmed exoplanets as in the ESI case.

\begin{table}[h!]
\centering
\caption{A sample of determined MSI} 
\begin{tabular}{l*{8}{c}r}
\hline
Names & Radius & Density &Temp & E. Vel & $MSI_S$ & $MSI_I$ & MSI\\ 
\ &  (MU) & (MU) & (K) & (MU) &  \ & \ &  \\\hline
Mars &	1.00 &	1.00 &	240 &	1.00  &	1.00 &	1.00 &	1.00\\
Earth &	1.88 &	1.407 &	288 &	2.23  &		0.659 &	0.70 &	0.68\\
Moon & 0.513 & 0.853 & 197 & 0.48 & 0.66 & 0.77 & 0.71\\
Kepler-42 d & 1.07 & 1.05  & 503& 1.11 & 0.58 & 0.95 & 0.74\\
Kepler-378 c & 1.29 & 1.08 & 462& 1.37 & 0.59 & 0.90 & 0.73\\
Kepler-438 b & 2.10 & 1.23  & 312& 2.35 & 0.70 & 0.73 & 0.72\\
Proxima Cen b& 2.10 & 1.23  & 259& 2.35 & 0.69 & 0.73 & 0.71\\
\hline
\end{tabular}
\label{Table:msi}
\end{table}

\subsection{ANALYSIS OF THE MSI}

\begin{figure}[h!]
\centering
\includegraphics[width=10cm,angle=0]{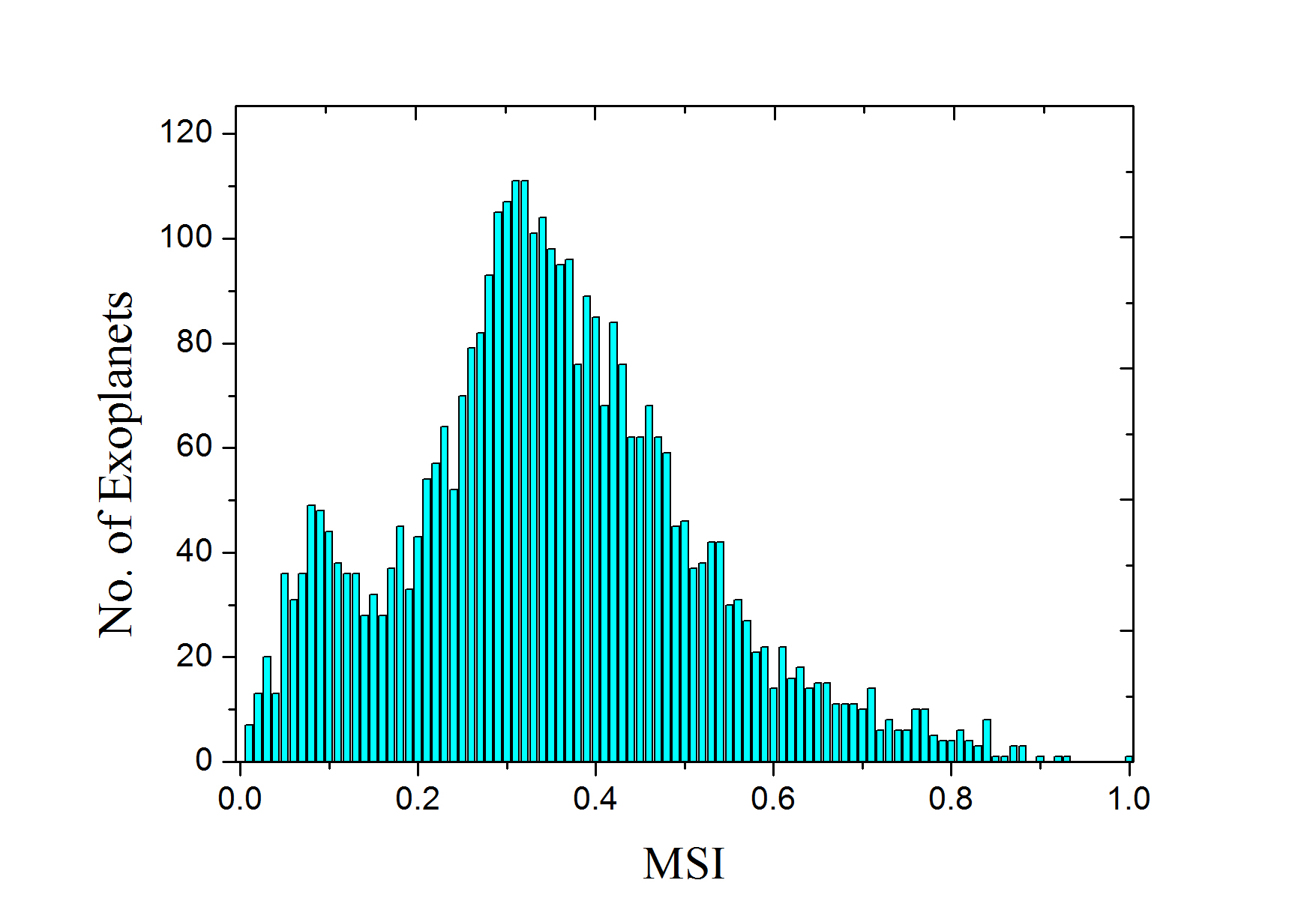}	
\caption{Histogram of the global MSI values for 3566 confirmed exoplanets.}
\label{fig:MSI}
\end{figure}

\begin{figure}[h!]
\centering
\includegraphics[width=10cm,angle=0]{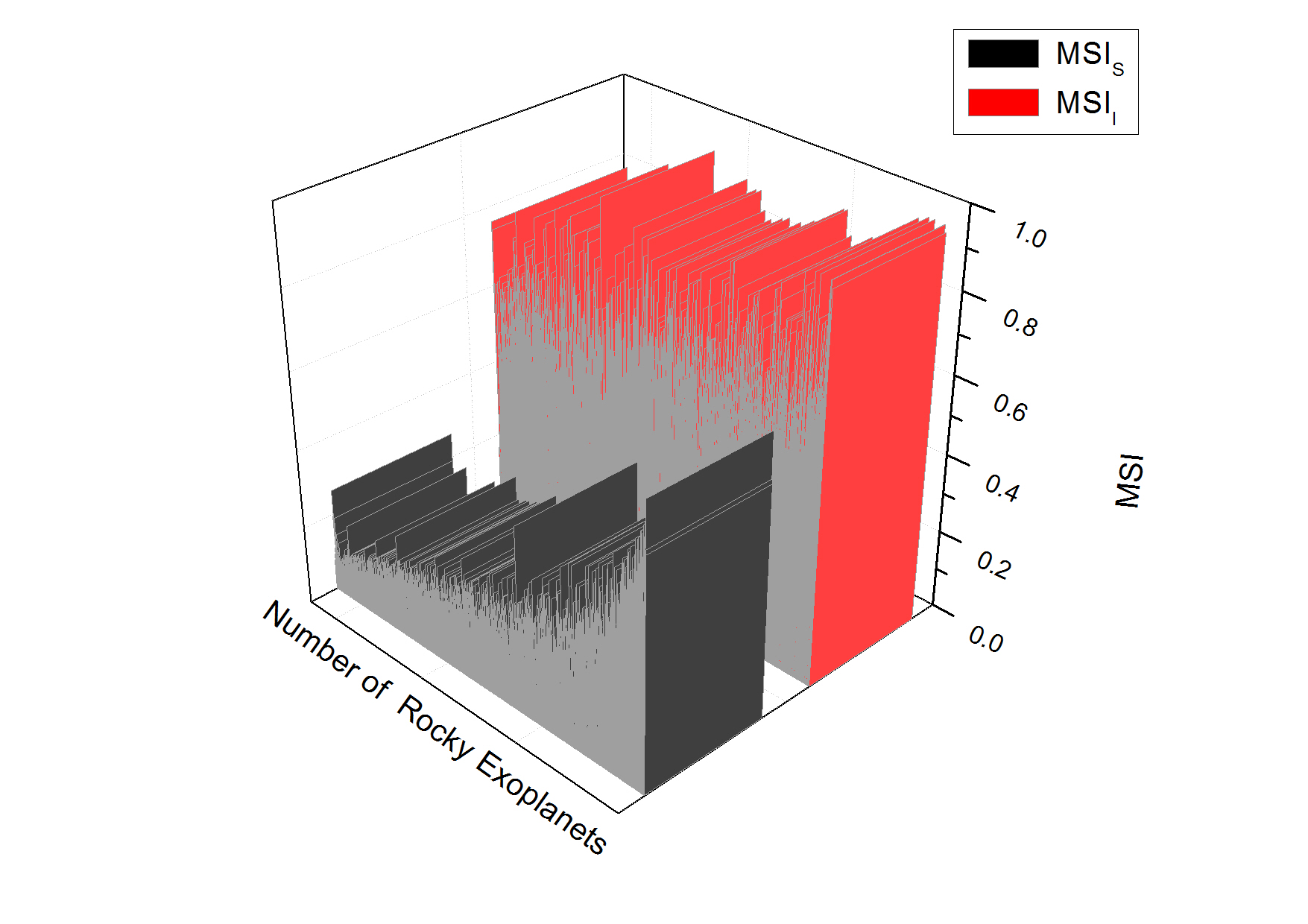} 
\caption{Interior and Surface MSI for 1650 rocky planets.} 
\label{fig:MSI_two}
\end{figure}

Results of Section~3.3 are represented as a histogram of the  global MSI (Fig.~\ref{fig:MSI}) for 3566 confirmed exoplanets. In Fig.~\ref{fig:MSI_two}, we show the 3-D histogram of the interior and surface MSI. As with the ESI, we can see that interior MSI is more dominant factor than surface MSI for the rocky exoplanets in the global MSI. 

\begin{figure}[h!]
\centering        
\includegraphics[width=13cm,angle=0]{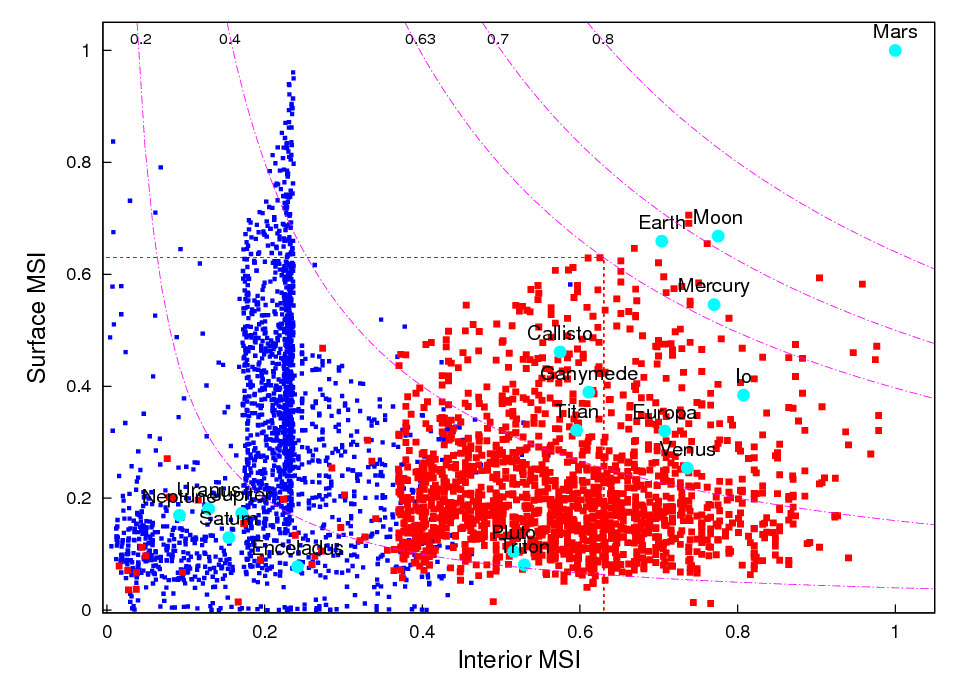} 
\caption{Plot of interior MSI versus surface MSI. Blue dots are the giant planets, red dots are the rocky planets, and cyan circles are the Solar System objects. The dashed curves are the isolines of constant global MSI, with values shown in the plot. Planets with MSI$\gtrsim 0.63$ are assumed Mars-like; we obtain 29 planets.}\label{figure: 3.7}
\label{fig:MSIplot}
\end{figure}

\begin{figure}[h!]
\centering
\includegraphics[width=8.7cm,angle=0]{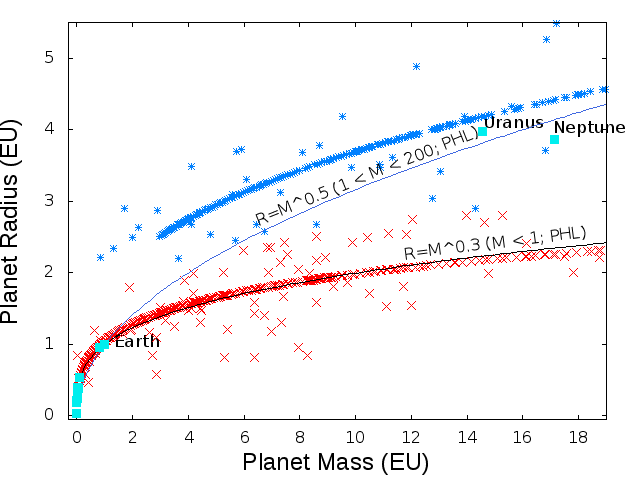}	
\includegraphics[width=8.7cm,angle=0]{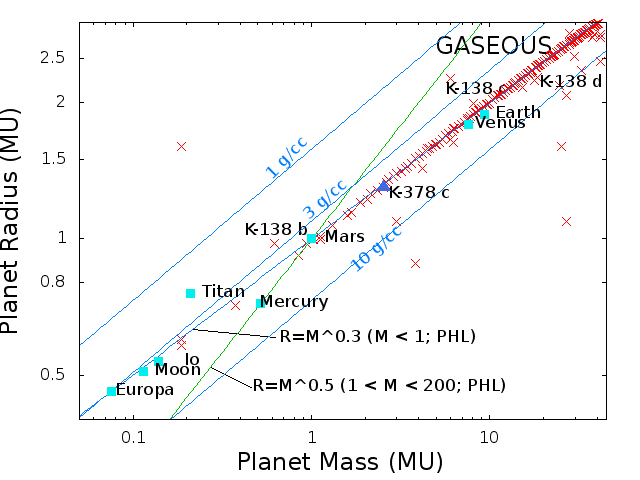}
\caption{{\it Left}: Mass-radius diagram for exoplanets with measured masses less than 20 EU along with model curves for different mass-radius relation: black line is $R=M^{0.3}$ for $M_{\rm E} < 1$; blue-dotted line is $R=M^{0.5}$ for $1< M_{\rm E}< 200$. Red crosses indicate rocky planets, blue crosses are gas giants, cyan squares are our Solar System objects. In 2016, the data in the catalog suggested only two rocky exoplanets smaller than
Earth. In the present data, there are many more smaller planets. {\it Right}: Blow-up of the previous plot for small-size planets, in terms of the Mars units. Line of same mass-radius relation are marked on the plot, along with the isolines of constant density. Some interesting planets are marked by names.}
\label{fig:density}
\end{figure}

In Fig.~\ref{fig:MSIplot}, we present a scatter plot of interior MSI versus surface MSI for 3566 confirmed exoplanets. The dashed curves are the isolines of constant global MSI, with values shown in the plot (the value for the Earth is 0.68). Planets above MSI $\sim 0.63$ are considered Mars-like. For example, Kepler-186 f has MSI$\gtrsim 0.69$ is potentially habitable for extremophile life forms. There is a noticeable difference of similarity to Mars planets (29) compared with the Earth-like planets, where we find 99 of the Earth-like PHPs. The reason is that, probably due to the selection bias, smaller planets are under-represented in the catalog of detected planets. In Fig.~\ref{fig:density}, we present mass-radius plots for small size planets. Left plot is in the Earth units for planets of $\leq 20$ Earth masses, and the right plot in Mars units for planets $\leq 2$ Earth  masses, or $\sim 20$ Mars masses. The regular features seen on this plot, as well as on Figs.~4 and~7, are due to the use of modelled values, inferring mass from radius or radius from mass. We have plotted the model curves for the mass-radius relation adopted by the PHL on both plots. It is obvious that data for the most of the rocky planets was estimated from these relations. Planets that are noticeably not following the relations are those whose mass and radius were estimated independently, such as, e.g. Kepler-138 b (Jontof-Hutter et al. 2015). 

\newpage 
\newpage
\section{Discussion and Conclusion} 

The current definition of the habitable planets as those that are to be found in habitable zones (HZ) of the hosts has a caveat. A habitable planet may not necessarily harbour life (e.g. Venus), although from a distance its biosignatures may show positive signs; while Earth-like planets (similar atmosphere, size, mass, at a similar distance and orbiting a solar-type star) with high probability may  actually have life on them (e.g. Jheeta 2013). Other factors can come in next, such as, for example, the age of the  planet \citep[e.g.][]{safonova} since life requires time to change its environment to become noticeable.

The search for habitable exoplanets has essentially two goals, both of which (if fulfilled) will have profound implications to our civilization. One is to seek life elsewhere outside the Earth. Another one is to seek a twin-Earth, preferably nearby. The second goal, in principle, is to have a planet habitable for our kind of life, but uninhabited, so that we can shift there in the far away future, if needed. Another aspect of the second goal is that it is easier to search for the biosignatures of life as we know it on a planet which looks just like Earth. It is estimated that one in five solar-type stars and approximately half of all M-dwarf stars may host an Earth-like planet in the HZ. Extrapolation of Kepler's data shows that in our Galaxy alone there could be as many as 40 billion such planets (e.g. Borucki et al. 2010; Batalha et al. 2013; Petigura et al. 2013). And it is quite possible that quite soon we may actually observe most of them. With the ultimate goal of a discovery of life, astronomers do not have millennia to quietly sit and sift through more information than even pentabytes of data. In addition, obtaining the spectra of an Earth-like planet around a Sun-like star is difficult, and would require a large-scale expensive space mission (such as e.g. JWST), which still may be able to observe only about a hundred stars over its lifetime (e.g., Turnbull et al. 2012). Thus, it is necessary to prioritise the planets to look at, to introduce a quick estimate of whether a planet can be habitable from the measured properties of the star and the planet. The Earth Similarity Index (ESI), a parametric index to analyze the exoplanets data, was introduced to access precisely that; to evaluate the potential habitability (Earth-likeness) of all discovered to date exoplanets. Since our search for habitable exoplanets (aka Earth-like life, which is clearly favoured by Earth-like conditions) is by necessity anthropocentric, as all we know for sure is only the Earth-based habitability, any such indexing has to be centred around finding Earth-like life, at least initially. Here, we have shown how the ESI can be derived from the initial mathematical concept of similarity.

Out of the four parameters, entering the global ESI, only one (radius) is a direct observable, while the remaining three parameters, surface temperature, escape velocity and density, are generally calculated. According to the PHL project, surface ESI is dominating the interior ESI, because the weight exponent value for the surface temperature is much higher than that of the interior parameters. We found, however, that the interior ESI is a predominant factor in the global ESI for the rocky exoplanets, where the real values of interior and surface ESI play a larger role than the weight exponents (Fig.~2). However, even though evaluation of only radius and density parameters may be enough to suggest a rocky nature of an exoplanet, due to the geometrical mean nature of the ESI formulation, we need to consider the surface temperature to verify the Earth-likeness. For example, if we consider surface temperature values as 10 K, 100 K and 2500 K, and keep interior ESI the same as for the Earth, the corresponding global ESI values will be 0.02, 0.40 and 0.11, respectively; clearly not habitable. Thus, the surface temperature and plays a key role in balancing the global ESI equation. However, there is always an observational difficulty in estimating the surface temperature value of the exoplanet. We introduced the calibration technique in Section 3.2 to try to mitigate this difficulty for the case of rocky planets. We have used the existing observational data from Solar System rocky objects and performed the linear regression to extrapolate to all rocky exoplanets. 

It is now believed that in its early history, Mars had a much wetter and warmer environment, just at the time when life on Earth is now known to have originated (this date was recently moved back to 4.1 Ga (Gigayears ago) (Bell et al. 2015). Curiosity data indicates early ($\sim 3.8$ Ga) Martian climate with stable water lakes on the surface for thousands to millions of years at a time (Grotzinger et al. 2015), and a recently discovered evidence of carbonate-rich ($\sim 3.8$ Ga) bedrock (Wray et al. 2016) suggested the habitable warm environment. It is possible that after the presumed catastrophic impact-caused loss of most of the atmosphere (e.g., Melosh et al., 1989; Webster et al. 2013), only the toughest life forms had survived, the ones we call here on Earth as extremophiles. They would have adapted to the currently existing conditions and just like the terrestrial extremophiles would need such conditions for the survival; for example, terrestrial methanogens have developed biological mechanism that allows them to repair DNA and protein damage to survive at temperatures from $-40^{\circ}$C to $145^{\circ}$C \citep{Tung2005}. The usual conditions for habitability would be different for such life forms. Carbon and water have the dominant role as the backbone molecule and a solvent of biochemistry for Earth life. However, the abundance of carbon may not be a useful indication of the habitability of an exoplanet. The Earth is actually significantly depleted in carbon compared with the outer Solar System. Here, on Earth, we have examples of life, both microbial and animal, that do not require large amounts of water either. For example, both bacteria and archaea are found thriving in the hot asphalt lakes (Schulze-Makuch et al. 2011b) with no oxygen and virtually no water present. They respire with the aid of metals, perhaps iron or manganese, and create their own water by breaking down hydrocarbons, just like {\it E. coli} gut bacteria that can generate most of their own water from light hydrocarbons (Kreuzer-Martin et al. 2005). We have introduced the Mars similarity index to study the Mars-like planets as potential planets to host extremophile life forms. In this scale, Moon has the MSI of $\sim 0.75$, Earth has the MSI of $\sim 0.68$, and the next closest exoplanet is Kepler-186 f (MSI=0.69), which is listed as potentially habitable planet in the HEC. Mars-like planets can tell us about the habitability of small worlds rather than planets that are far from their star. For example, Earth at Mars distance would most probably still be habitable \citep{MarsClimate}. Given constant exchange of impact ejecta between the planets, it is possible that biota from the Earth reached and survived on Mars, which thus could have been `extremophile'-habitable throughout all its history. It is interesting to note that when we started this work, only two small (less than Earth in size or mass) exoplanets were known. In this year, many new small planets were discovered, with resulting 29 planets that we can call Mars-like.

So can such similarity scale be useful? It actually might be developed into a sort of same scale as the stellar types in astronomy, such as classifying stars based on information about size, temperature, and brightness. It can be used as a quick tool of screening planets in important characteristics in Earth-likeness. Different ranking scales for evaluating habitability perspectives for follow-up targets have been already proposed (e.g. habitability index for transiting exoplanets (HITE, Barnes et al. 2015), or Cobb-Douglas Habitability Index (CDHI, Bora et al. 2016). We conclude that it is necessary to arrive at the a multiparameter calculator that, though based on a current similarity scale -- ESI, may include more input parameters (e.g. orbital properties, temperature, escape velocity, radius, density, activation energy and so on), and extended to other planet-likeness, such as similarity to Mars -- MSI. If we find habitable possibilities on Titan for example, the scale can be modified for the Titan-like habitability. We would like to call this future calculator a Life Information Score (LIS), which shall be used as an overall calculator to detect life itself. The LIS is almost similar to the anthropic selection, which basically deals with the preconditions for the emergence of life and, ultimately, intelligent observers \citep{Wal}. But the expected outcome of this LIS is to accurately measure the possibility of a planet to host any form of life using only the parametric data. 

\appendix
\section*{Calculation of Mars ESI as an example}

$ESI_x$ calculations for Mars are performed using Eq.~(\ref{eq:esi}), with weight exponents from Table~\ref{table:3.1}, by using the following values for the input parameters, 
\begin{align}
&R=0.53 \times 6371\,km = 3376.63\,km\,,\nonumber\\
& \rho=0.71 \times 5.51\, g/cm^3 = 3.9121 \,g/cm^3\,,\nonumber\\
&V_{e}= 0.45 \times 11.19\, km/s = 5.0355\, km/s \,,\nonumber\\
&T_{s}= 240 \, K\,. \nonumber
\end{align}
The ESI for each parameter are, accordingly,
\begin{align}
&ESI_R=\left(1-\left|3376.63\,km-6371\,km\right|\Big/ \left|3376.63\,km +6371\,km\right|\right)^{0.57} = 0.8124\,,\nonumber\\ 
& ESI_{\rho} =\left(1-|3.9121\, g/cm^3 - 5.51 \,g/cm^3|\Big/|3.9121 \,g/cm^3+5.51 g/cm^3|\right)^{1.07} = 0.8218 \,,\nonumber \\
& ESI_{v_{e}} =\left(1-|5.0355\, km/s-11.19\, km/s|\Big/|5.0355\, km/s+11.19\, km/s|\right)^{0.7} = 0.7162\,,\nonumber\\ 
&ESI_{T_{s}} =\left(1-|240\,K-288\,K|\Big/|240\,K+288\,K|\right)^{5.58} = 0.5875\,.\nonumber
\end{align} 
Interior ESI from Eq.~(\ref{eq:interiorESI}) is:\\
 
$ESI_I = \sqrt{0.8124\times0.8218} \approx 0.8171$.\\

\noindent
Surface ESI from Eq.~(\ref{eq:surfaceESI}) is:\\

$ESI_S = \sqrt{0.7162 \times 0.5875} 
 \approx 0.6487$.\\
 
\noindent 
And the global ESI for Mars (Eq.~\ref{eq:globalESI}) is:\\
 
 $ESI = \sqrt{0.8171 \times 0.6487} \approx 0.728$.

\section*{Acknowledgments}

We would like to thank Jayant Murthy (Indian Institute of Astrophysics, Bangalore) and Yuri Shchekinov (Lebedev Institute, Moscow) for the fruitful discussions. This research has made use of the Extrasolar Planets 
Encyclopaedia at {\tt http://www.exoplanet.eu}, Exoplanets Data Explorer at {\tt http://exoplanets.org}, the Habitable Zone Gallery at {\tt http://www.hzgallery.org/}, the NASA Exoplanet Archive, which is operated by the California Institute of Technology, under contract with the National Aeronautics and Space Administration under the Exoplanet Exploration Program at {\tt http://exoplanetarchive.ipac.caltech.edu/} and NASA Exoplanet Archive at {\tt http://exoplanetarchive.ipac.caltech.edu} and NASA Astrophysics Data System Abstract Service.

\end{document}